# Enhancing Green Economy with Artificial Intelligence: Role of Energy Use and FDI in the United States

Abdullah Al Abrar Chowdhury[1], Azizul Hakim Rafi[1], Adita Sultana[1], Abdulla All Noman[2]

[1]Information Technology of Science, American National University, 1813 East Main Street, Salem, VA 24153.
[2]Montclair State University, Montclair, NJ, USA 07043.

Corresponding author: Azizul Hakim Rafi, Email: rafiazizul96@gmail.com,

**Abstract**
The escalating challenge of climate change necessitates an urgent exploration of factors influencing carbon emissions. This study contributes to the discourse by examining the interplay of technological, economic, and demographic factors on environmental sustainability. This study investigates the impact of artificial intelligence (AI) innovation, economic growth, foreign direct investment (FDI), energy consumption, and urbanization on $CO_2$ emissions in the United States from 1990 to 2022. Employing the ARDL framework integrated with the STIRPAT model, the findings reveal a dual narrative: while AI innovation mitigates environmental stress, economic growth, energy use, FDI, and urbanization exacerbate environmental degradation. Unit root tests (ADF, PP, and DF-GLS) confirm mixed integration levels among variables, and the ARDL bounds test establishes long-term co-integration. The analysis highlights that AI innovation positively correlates with $CO_2$ reduction when environmental safeguards are in place, whereas GDP growth, energy consumption, FDI, and urbanization intensify $CO_2$ emissions. Robustness checks using FMOLS, DOLS, and CCR validate the ARDL findings. Additionally, Pairwise Granger causality tests reveal significant one-way causal links between $CO_2$ emissions and economic growth, AI innovation, energy use, FDI, and urbanization. These relationships emphasize the critical role of AI-driven technological advancements, sustainable investments, and green energy in fostering ecological sustainability. The study suggests policy measures such as encouraging green FDI, advancing AI technologies, adopting sustainable energy practices, and implementing eco-friendly urban development to promote sustainable growth in the USA.

## Introduction

The environment is crucial for supporting life on Earth; nevertheless, the increasing global pollution and waste have emerged as a significant problem in recent years [1,2]. The majority of climate-altering and planet-warming substances emitted into the atmosphere are carbon dioxide ($CO_2$) [3]. The United States (USA) distinguishes itself among these nations through its evolution as a dominant force in consumption as well as production [4]. In 2020, the USA emitted 5,416 metric tons (MT) of $CO_2$, or almost 16% of global emissions [5]. However, in 2021, the United States reinforced its dedication to the Paris Agreement, adopting an ambitious Nationally Determined Contribution to reach a 50–52% drop in net greenhouse gas (GHG) emissions by 2030 [6,7]. The United States consumes the largest quantity of fossil fuel annually, totaling 913.3 million tons of oil, which is 50% greater than China's consumption, the second highest globally [8,9]. Alleviating the adverse effects of global climate change has emerged as a worldwide priority, with a crucial component of this effort being the reduction of $CO_2$ emissions [10,11]. In accordance with the UN 2030 universal sustainable development objective, the advancement of renewable energy is prioritized above everything else. Consequently, the USA has considerable accountability for the climate crisis and global warming, being one of the foremost emitters of GHGs; thus, assessing its environmental sustainability is a critical issue. In this context, our research question is to examine the response of GDP, FDI, energy consumption, urbanization, and AI innovation on $CO_2$ emissions in the context of United States.

Several scholars have thoroughly investigated the ecological conditions in the United States, mostly utilizing pollution metrics such as $CO_2$ emissions [12,13]. By encouraging domestic investment, facilitating technological transfers in the receiving nation, and boosting the development of human capital, foreign direct investment (FDI) enhances economic growth and is therefore essential to economic development [14]. Another viewpoint holds that FDI improves host countries' surroundings by bringing cutting-edge environmental technologies and sustainable practices [15,16]. The environment field is not an exception to the dramatic transformations we have lately experienced in numerous industries as artificial intelligence (AI) has grown more integrated. The United States leads in AI development and will be essential in developing suitable guardrails and regulatory frameworks that promote the responsible use of the technology [17]. The development of AI technologies can help address several global green growth concerns. Furthermore, applying AI can reduce environmental emissions [18]. Businesses are facing pressure to curtail GHG emissions and promote environmentally friendly practices as the global community battles climate change.

In 2020, the United States was the second-largest contributor to world pollution, producing 4.7 billion metric tons of $CO_2$. Multiple studies has demonstrated the mixed consequences of GDP on $CO_2$ emission in different regions [19,20,21]. After acknowledging this fact, politicians and governments must create plans for resource sustainability and seek solutions that strike a balance between ecological sustainability and GDP [22]. Increased use of clean energy sources is thought to be able to lessen the negative economic effects of climate change [23,24]. Between 2005 and 2022, emissions from the energy sector decreased by 35.9%, during which the sector represented almost one quarter of total US emissions. Furthermore, the United States' share of GDP in 2015 that went toward renewable energy was 0.2%, far less than that of other developing nations like South Africa (1.4%), China (0.9%), India (0.5%), and Brazil (0.4%) [25]. With a 15.8% share of primary energy consumption and 13.8% share of $CO_2$ emissions in 2020, the United States is also the world's top primary energy consumer [26].

This study enhances existing research in the following manners: Firstly this paper critically analyzes the correlation among AI innovation, foreign direct investment, and $CO_2$ emissions in the United States. Secondly, unlike previous studies that solely examined developing nations, this research focuses on one of the world's most industrialized countries with a robust financial system, with the aim of investigating and clarifying the relationship between financial systems and environmental sustainability. Moreover, it is the inaugural investigation of the dynamic interconnections among AI innovation, energy consumption, and foreign direct investment (FDI). Specifically the United States provides an ideal context for investigating the correlation between economic activity and environmental sustainability, given its significant ecological deficit and developed economy. This study reveals that while AI innovation reduces $CO_2$ emissions, factors such as GDP, FDI, urbanization, and energy consumption exacerbate environmental degradation. This unique contribution is vital for stakeholders as it offers tangible insights into how industrialized countries, like the USA, may harmonize economic growth with environmental sustainability. Additionally, the research corroborates the STIRPAT framework along with sophisticated methods like ARDL in the United States, employing a contemporary dataset from 1990 to 2022. Furthermore, it employs a comprehensive approach and innovative econometric models to offer significant insights that aid the United States in its endeavors to attain SDG-7 and SDG-13, especially in the quest for carbon neutrality. This research offers governments, corporations, and environmental advocates scientifically substantiated ideas for reconciling economic and environmental objectives, chiefly via financial, technological, and industrial modifications.

The subsequent portions of this work are organized as follows: Section II examines the pertinent literature, Section III delineates the methodology and data, Section IV showcases the results and analysis, and Section V explores the policy implications and conclusions.

## Literature Review
It is clear from a thorough analysis of earlier research that not many studies have looked into the connection between GDP, energy use, FDI, $CO_2$ emissions, and AI innovation. The bulk of publications concentrated on the effects of trade openness, urbanization, and green energy usage on environmental quality, even though several investigations looked at the ARDL model. The USA hasn't seen much in-depth research on ecological degradation, a relatively young discipline. Nevertheless, the research drew upon a few previous studies to guide the selection of variables and methods. This section will address a select few of these questions.

### *GDP and $CO_2$ emission*
Numerous researches have been conducted to ascertain the correlation between environmental and economic activity. For example, Manu and Sulaiman [27] examined the impact of energy consumption and economic growth on Malaysia's $CO_2$ emissions utilizing the OLS method for the years 1965 to 2015. Their research demonstrated that $CO_2$ emissions diminish as revenue escalates. The favorable relationship between economic expansion and CO2 emission is found by Saudi et al.[28], and Zubair et al.[29]. Conversely, Sarkar et al. [30] analyzed time-series data from 1980 to 2016 to investigate the relationship among Malaysia's energy use, $CO_2$ emissions, and economic expansion. The empirical evidence indicated that usage of energy and economic expansion substantially elevated $CO_2$ emissions in Malaysia from 1980 to 2016. In order to examine the relationship between GDP and $CO_2$ emissions, Chen et al. [31] used China's annual data from 1990 to 2020 and the QARDL technique. They discovered that China's GDP has a positive effect on $CO_2$ emissions. Etokakpan et al. [32] analyzed the relationship between capital formation, globalization, $CO_2$ emissions, and GDP in Malaysia using a dataset from 1980 to 2014 within a multivariate framework. The authors employed an innovative combined co-integration test to ascertain the magnitude of the long-run equilibrium relationship. The empirical findings indicated that GDP adversely affected environmental quality. Multiple researcher such as Wada et al.[33] within Brazil, Rjoub et al. [34] in Turkey and He at al.[35] within Mexico found the same conclusions. On the other hand, Muhammad et al. [36] employed FMOLS and two-stage least squares regression methods to analyze the correlation between GDP growth and $CO_2$ emissions. The findings corroborated the U-shaped link indicated by the EKC theory in high- and upper-middle-income nations.

### *AI innovation and $CO_2$ emission*
The loss of biodiversity and global warming are complex concerns that require advanced and inventive solutions [37]. Environmental professionals anticipate several benefits from AI tools [38]. Policymakers can use AI innovation to develop scientifically supported plans and strategies for green ecosystems [39]. Existing research indicates that investigations on carbon reduction related to AI are in their infancy. A pertinent study by Liu et al. [40] analyzes industrial robot data from 16 sectors in China from 2006 to 2016 to investigate the correlation between AI and energy intensity. The implementation of AI technology in the industrial sector diminishes energy intensity by enhancing industrial output while decreasing energy consumption and environmental degradation. Chen et al. [41] investigate the impact of AI on carbon emissions utilizing panel data from 270 Chinese cities spanning 2011 to 2017. Their empirical findings indicate that AI decreases carbon emissions by optimizing production processes, boosting communication facilities, and advancing green technological innovation. Negi [42] examines the investment trends in artificial intelligence originating from the three leading nations: China, India, and the United States. The paper delineates the measures the government has implemented to integrate artificial intelligence into its existing ecosystem. Green AI can enhance productivity and mitigate its adverse environmental impacts [43].

### *Energy Use and $CO_2$ emission*
The primary contributor to climate change is the combustion of fossil fuels, which releases significant quantities of greenhouse gases into the environment [44]. But long-term cost reductions from using alternative or green energy sources will raise people's overall level of living [45]. Raihan et al. [46] examine the relationship between energy use and $CO_2$ emissions in Malaysia from 1990 to 2019. They employed the ARDL and DOLS methodologies, which revealed a positive and significant energy use coefficient in relation to $CO_2$ emissions, indicating a 0.91% rise in $CO_2$ emissions for every 1% increase in energy usage. Adebayo and Kalmaz [47] employed ARDL, FMOLS, and DOLS estimators to reveal a substantial positive correlation between power consumption and $CO_2$ emissions in Egypt, utilizing data from 1971 to 2014. Moreover, Adebayo et al. [48] identified a positive correlation between $CO_2$ emissions and energy consumption by employing the ARDL model for MINT nations, covering the period from 1980 to 2018. Conversely, Raihan and Tuspekova [49] examined the correlation between green energy utilization and CO2 emissions in Peru from 1990 to 2018. Using the DOLS and ARDL methodologies, they found a negative correlation and statistical significance in the utilization of clean energy, suggesting that a 1% increase in green power use results in a 0.52% reduction in CO2 emissions over the long term. Multiple researcher such as Abbasi and Adedoyin [50] in China, Balsalobre-Lorente et al.[51] across BRICS countries, Saqib [52] in MENA region corroborated the destructive effect of energy usage on environment quality. On the other hand, Younis et al.[53], Salari et al.[54] concluded that energy consumption boost the environment quality. In the end, energy-efficiency regulations can enhance energy conservation, while increased investment in energy production and the promotion of energy savings will diminish carbon emissions [55].

*FDI and $CO_2$ emission*
Foreign Direct Investment (FDI) can facilitate the transfer of cleaner technologies and sustainable practices, leading to enduring reductions in emissions [29,56]. Jafri et al. [57] investigate the asymmetric impact of FDI on $CO_2$ emissions utilizing the NARDL methodology for China from 1981 to 2019. Their findings indicate that a positive shift in FDI has a comparatively greater impact on $CO_2$ emissions. Moreover, several studies also found the same conclusion in different region [58,59,60,61]. On the other hand, Lin et al. [62] examine the impact of FDI on emission reduction in China from 2004 to 2015, employing geographic Durbin economic models with two-way fixed effects. The findings indicate that FDI facilitates a decrease in emissions nationwide. Furthermore, Eskeland and Harrison [63] contended that FDI typically accompanies energy-efficient technologies and may positively impact the natural world. Similar outcomes were also observed by Wang et al. [64] in China, Pata and Samour [65] within France, and Abbas et al.[66] in developing countries. Conversely, Haug and Ucal [67] shown that spikes in FDI had no statistically significant long-term effects on per capita CO2 emissions.

*Urbanization and $CO_2$ emission*
Economic growth and industrialization-driven urbanization are contributing to the rising utilization of fuels that generate greenhouse gas emissions [68].The concluding section of the research examines previous studies on the empirical relationship between urbanization and $CO_2$ emission. For instance, Mahmood et al. [69] examine the impact of urbanization on per capita $CO_2$ emissions in Saudi Arabia, analyzing data from the years 1968 to 2014. The findings indicate that urbanization hinders the environment due to its elastic impact on emissions. Parshall et al. [70] examined the substantial impact of urbanization on the circular economy and environmental health in the USA. Raihan et al. [71] explore the impact of urbanization on the load capacity factor in Mexico from 1971 to 2018. This study utilizes the ARDL approach and demonstrates that urbanization decreases Mexico's LCF, thereby degrading the ecology. Sufyanullah et al. [72] examined the impact of urbanization on $CO_2$ emissions in Pakistan. They employed the ARDL model and noted that $CO_2$ emissions rise with increased urbanization. Additional studies [73,74,75,76,77] suggest that urbanization increases $CO_2$ emissions in the atmosphere. In comparison, Xu et al. [78] evaluated the impact of urbanization on the ecosystem in Brazil from 1970 to 2017. Unexpectedly, the results of the ARDL methodology indicated that urbanization does not influence the surroundings in Brazil. Moreover, Haseeb et al. [79] revealed analogous findings utilizing FMOLS from 1995 to 2014, suggesting that URB had no significant impact on environmental quality in the BRICS nations. Nonetheless, as highlighted by Martinez et al. [80] urbanization may contribute to addressing climate change due to the heightened knowledge among these groups. Moreover, Diputra and Baek [81] determined that urbanization exerted no substantial impact on emissions in Indonesia.

*Literature Gap*
The current status of research indicates that studies on the USA are scarce. The empirical literature lacks an evaluation of the geographical implications of macroeconomic variables on $CO_2$ emissions in the USA. The literature on the STIRPAT framework and ARDL methodology is deficient in the USA and similarly limited in other global locations. This study tries to fill in that gap by looking at modern variables using advanced methods from a U.S. point of view. The goal is to find the main and secondary effects of GDP, AI innovation, energy consumption, urbanization, and FDI on $CO_2$ emissions in this area. By examining these procedures, the USA may determine whether leveraging technological innovation, financial integration, and commercial expansion can enhance its ecosystem quality and align it with global trends toward greater environmental sustainability.

## Methodology
*Data and Variables*
This study seeks to observe the impact of GDP, FDI, AI innovation, energy use, and urbanization on $CO_2$ emissions in the USA. The study utilizes $CO_2$ emissions as an indicator of ecological health, employing data from the World Development Indicators (WDI) database as the dependent variable. Statistics on AI innovation are sourced from Our World in Data, while data for GDP, energy consumption, foreign direct investment (FDI), and urbanization are derived from the revised WDI. These variables encompass annual data from 1995 to 2022. Table 1 presents a complete enumeration of each variable along with their respective details and a sign chosen for this research.

Table 1. Variables description

| Variables | Description | Logarithmic Form | Unit of Measurement | Source |
|---|---|---|---|---|
| $CO_2$ | $CO_2$ Emission | $LCO_2$ | $CO_2$ Emission (kt) | WDI |
| GDP | Gross Domestic Product | LGDP | GDP per capita (current US$) | WDI |
| AI | AI Innovation | LPAI | Estimated Investment in AI (US$) | Our World in Data |
| ENU | Energy use | LENU | Energy use (kg of oil equivalent per capita) | WDI |
| FDI | Foreign Direct Investment | LFDI | Net Inflows (Current US$) | WDI |
| URB | Population | LPOP | Population, total | WDI |

*Theoretical Framework*
The STIRPAT paradigm holds substantial importance for environmental research. This method is a versatile analytical tool that facilitates the understanding of intricate relationships between human societies and the environment, irrespective of the subject matter, including GHG emissions, air pollution, deforestation, or biodiversity loss [82]. By using STIRPAT, this study allows for an extensive evaluation of numerous variables that affect $CO_2$ emissions in the context of USA. The IPAT model was proposed by Ehrlich and Holdren [83] and is phrased as follows:

$$I = \int PAT \qquad (1)$$

Nonetheless, this model proved challenging to evaluate hypothetically [84]. To address these constraints, Dietz and Rosa [85] expanded the IPAT model, resulting in the STIRPAT model. This research can effectively assess the interplay between human activities and environmental outcomes by representing $CO_2$ emissions as the 'I' or impact variable, urbanization as the 'P', and other variables like AI innovation, GDP, energy consumption, as the 'T' or technological variables. Equation (2) shows the functional form of the STIRPAT framework:

$$I_i = C \cdot P_i^{\alpha} \cdot A_i^{\beta} \cdot T_i^{\gamma} \cdot \varepsilon_i \qquad (2)$$

Based on a comprehensive review of relevant literature, the empirical model employed in this work yielded the subsequent approximations.

$$Environmental\ Impact = f(Population, Affluence, Technology) \qquad (3)$$

To assess the effects on the environment, this study uses $CO_2$ emissions as a proxy indicator. This is the expression that may be used to derive Equation (4):

$$CO_{2it} = \delta_0 + \delta_1 GDP_{it} + \delta_2 AI_{it} + \delta_3 ENU_{it} + \delta_4 FDI_{it} + \delta_5 URBA_{it} + \varepsilon_{it} \qquad (4)$$

Here, GDP stands for gross domestic product; AI innovation was represented by AI, energy use through ENU, foreign direct investment via FDI and urbanization by URBA. Equation (5) makes use of the logarithmic transformation of variables to ensure that the information has a normal distribution. By employing the logarithmic structure, the info is standardized which makes it more consistent with the assumptions that underlie numerous statistical methods.

$$LCO_{2it} = \delta_0 + \delta_1 LGDP_{it} + \delta_2 LAI_{it} + \delta_3 LENU_{it} + \delta_4 LFDI_{it} + \delta_5 LURBA_{it} + \varepsilon_{it} \qquad (5)$$

*Empirical Methods*
This study utilized the ARDL approach for data analysis to explore the correlation between $CO_2$ emissions and several independent variables in the USA. To ensure robustness, the Fully Modified Ordinary Least Squares (FMOLS), Dynamic Ordinary Least Squares (DOLS), and Canonical Cointegration Regression (CCR) methods were additionally employed. Initially, the author performed unit root tests (ADF, P-P, and DF-GLS) to verify stationarity. The properties of the time series data led to the employment of the ARDL limits test. Subsequently, both short-term and long-term ARDL estimates were derived, followed by the Pairwise Granger causality examination. Ultimately, multiple diagnostic assessments were conducted, allowing us to find the most precise and effective econometric model following a comprehensive evaluation procedure.

*Unit root test*
Performing a unit root testing is needed to avert erroneous regression analysis. This test can determine the degree of integration [86]. The Dickey-Fuller Generalized Least Squares, [87] Phillips-Perron [88], and Augmented Dickey-Fuller [89] unit root tests were used to see if the data set was stationary. In comparison to the Dickey-Fuller (DF) method, the Augmented Dickey-Fuller (ADF) technique is more resilient and suitable for more complex procedures [90]. The DF-GLS test shows superior overall performance regarding small sample size and power, surpassing the conventional Dickey-Fuller test [91]. Before employing the ARDL bound test estimator for cointegration analysis, it is essential to conduct the ADF and PP unit root tests, as the estimator is applicable only when the variables are stationary at the level or first difference [92].

*ARDL technique*
The ARDL technique, which includes the features of both distributed lag and autoregressive models, was proposed by Pesaran et al. [93]. It is a comprehensive dynamic regression model that offers various advantages over traditional cointegration methods. Firstly, the model allows for the integration of variables to multiple orders, including order one, order zero, and even fractional integration, with the exception of 1(2). Moreover, unlike previous cointegration approaches that necessitated prior identification of a series' integration characteristic, this approach does not mandate any preliminary validation [94]. Due to its efficiency, we can employ this methodology for data analysis in scenarios with limited and small sample sizes [95,96]. Equation (6) can be used to represent the ARDL bound test:

$$\Delta LCO_{2t} = \beta_0 + \beta_1 LCO_{2t-1} + \beta_2 LGDP_{t-1} + \beta_3 LAI_{t-1} + \beta_4 LENU_{t-1} + \beta_5 LFDI_{t-1} \\ + \beta_6 LURBA_{t-1} + \sum_{i=1}^{m} \alpha_1 \Delta LCO_{2t-i} + \sum_{i=1}^{m} \alpha_2 \Delta LGDP_{t-i} + \sum_{i=1}^{m} \alpha_3 \Delta LAI_{t-i} \\ + \sum_{i=1}^{m} \alpha_4 \Delta LENU_{t-i} + \sum_{i=1}^{m} \alpha_5 \Delta LFDI_{t-i} + \sum_{i=1}^{m} \alpha_6 \Delta LURBA_{t-i} + \varepsilon_t$$

(6)

Where, m is the optimum lag length. Pesaran et al. [93] suggest using critical values for both upper and lower bounds to compare F-statistics. When the F-statistic surpasses the upper critical value, we reject the null hypothesis (H0), indicating a persistent relationship. If the F-statistic stays below the crucial value, we retain the null hypothesis (H0). The long-run coefficient estimate is derived from equation (7), which also validates the cointegration of the parameters. It uses an approximation of the Error Correction Term (ECT) to figure out short-term dynamic parameters based on long-term estimates [97]. The ECT is built into the ARDL structure. The equation (7) outlines the ARDL long-run equation presented below.

$$\Delta LCO_{2t} = \beta_0 + \beta_1 LCO_{2t-1} + \beta_2 LGDP_{t-1} + \beta_3 LAI_{t-1} + \beta_4 LENU_{t-1} + \beta_5 LFDI_{t-1} \\ + \beta_6 LURBA_{t-1} + \sum_{i=1}^{m} \alpha_1 \Delta LCO_{2t-i} + \sum_{i=1}^{m} \alpha_2 \Delta LGDP_{t-i} + \sum_{i=1}^{m} \alpha_3 \Delta LAI_{t-i} \\ + \sum_{i=1}^{m} \alpha_4 \Delta LENU_{t-i} + \sum_{i=1}^{m} \alpha_5 \Delta LFDI_{t-i} + \sum_{i=1}^{m} \alpha_6 \Delta LURBA_{t-i} \\ + \Omega ECT_{t-1} + \varepsilon_t \quad (7)$$

*Robustness Check*
To evaluate the precision of ARDL outcomes, we utilized the FMOLS, DOLS, and CCR techniques. The FMOLS approach is utilized to analyze a singular cointegrating relationship involving a combination of integrated orders of I(1) variables. The primary objective of this method is variable transformation. Phillips and Hansen [98] say that the FMOLS method fixes the problems with traditional cointegration methods that make it hard to draw conclusions. This makes the estimated t-statistics for long-term estimates more reliable. The DOLS technique may assist in integrating individual variables within a cointegrated framework when faced with a mixed order of integration. The dependent variable is calculated utilizing levels, leads, and lags as explanatory variables [99]. Nevertheless, as emphasized by Pesaran [100], the principal advantage of the DOLS prediction is its allowance for the varying order integration of distinct components within the cointegrated framework. Additionally, Park [101] proposed the CCR technique to examine cointegrating vectors within a model characterized by an integrated process of order I(1). The model's characteristics exhibit a significant similarity to FMOLS.

*Pairwise granser causality*
The study employed the paired Granger-causality test, devised by Granger [102], to ascertain the existence of a causal relationship among the factors. We can use F tests to assess Granger causality between variables X and Y, and the OLS test for coefficient estimation. The symbols Xt and Yt denote the values of the variables at time t, illustrating the time series for this variable pair. A bivariate autoregressive model may exhibit the variables Xt and Yt.

$$X_t = \beta_1 + \sum_{i=1}^{n} \alpha_i Y_{t-i} + \sum_{i=1}^{n} \mu_i X_{t-1} + e_t \qquad (8)$$
$$Y_t = \beta_2 + \sum_{i=1}^{n} \Omega_i Y_{t-1} + \sum_{i=1}^{n} \infty_i X_{t-i} + u_t \qquad (9)$$

Here, the information criterion determines the "n" number of lags. The parameters used for the assessment were $\beta_1$, $\beta_2$, $\alpha_i$, $\Omega_i$, $\mu_i$, and $\infty_i$.

*Diagnostic test*
The errors in Equation (7) must not exhibit serial correlation. This study employed various diagnostic techniques to verify the normality, serial correlation, and heteroscedasticity of the data. Three tests are needed in time series analysis to make sure that model assumptions are correct and that results are stable: the Lagrange Multiplier (LM) test, the Jarque-Bera test [103], and the Breusch-Pagan-Godfrey test [104]. The Jarque-Bera test assesses the normality of residuals, a crucial step since many econometric models require normally distributed errors for precise inference. The Lagrange multiplier test examines residuals for serial correlation to verify that errors do not correlate with time, thereby preventing biased and misleading estimates. The Breusch-Pagan-Godfrey test can yield inaccurate estimates and standard errors due to heteroscedasticity, or the non-constant variance of residuals. The model's stability was assessed by CUSUMSQ studies [105].

*Results and Discussion*
Table 2 lays out the descriptive statistics for the considered variables, derived from 32 observations. The table provides the mean, standard deviation, minimum, and maximum values for six variables in the USA: LCO2, LGDP, LAI, LENU, LFDI, and LURBA. All examined variables demonstrate positive mean values, with LCO2 showing the highest mean. LFDI has a minimum value of 2.268, while LURBA has the highest number.

Table 2. Summary Statistics.

| Variable | Obs | Mean | Std. Dev. | Min | Max |
| --- | --- | --- | --- | --- | --- |
| T | 32 | 2005.5 | 9.381 | 1990 | 2021 |
| LCO2 | 32 | 15.464 | .08 | 15.279 | 15.569 |
| LGDP | 32 | 10.644 | .319 | 10.081 | 11.159 |
| LAI | 32 | 7.506 | 1.036 | 6.321 | 9.724 |
| LENU | 32 | 4.77 | .322 | 3.949 | 5.272 |
| LFDI | 32 | 2.625 | .133 | 2.268 | 2.871 |
| LURBA | 32 | 19.5 | .087 | 19.335 | 19.621 |

Moreover, all variables have relatively small standard deviations, indicating a tight clustering of data points around the mean with limited temporal variability. Table 3 presents the results of all three stationarity tests (ADF, DF-GLS, and P-P) for log-transformed data in both level I(0) and first-difference I(1) forms. In all three unit root assessments, it seems that only the urbanization variable is stationary at level I(0), whereas $CO_2$, GDP, AI innovation, energy consumption, and FDI were non-stationary prior to examining their initial differences. This mixed sequence of integration prompts us to proceed with the assessment immediately, using the ARDL methodology.

Table 3. Results of unit root tes.t

| Variables | ADF | | P-P | | DF-GLS | | Decision |
|---|---|---|---|---|---|---|---|
| | I(0) | I(1) | I(0) | I(1) | I(0) | I(1) | |
| $LCO_2$ | -0.233 | -3.941*** | -0.231 | -4.001*** | -0.234 | -3.991*** | I(1) |
| LGDP | -0.872 | -4.091*** | -0.782 | -4.891*** | -0.809 | -4.091*** | I(1) |
| LAI | -0.704 | -5.105*** | -0.802 | -5.323*** | -0.756 | -5.034*** | I(1) |
| LENU | -0.172 | -5.011*** | -0.177 | -5.071*** | -0.819 | -2.150*** | I(1) |
| LFDI | -0.072 | -4.108*** | -0.065 | -4.015*** | -0.025 | -4.342*** | I(1) |
| LURBA | -5.012*** | -7.011*** | -5.801*** | -7.605*** | -5.831*** | -7.050*** | I(0) |

Standard errors in parentheses
*** $p<0.01$, ** $p<0.05$, * $p<0.1$

This research employed the ARDL bounds testing approach to ascertain the presence of co-integration among the variables. The F-statistic of 5.3421 is more than the critical value, indicating that the null hypothesis of no co-integration is rejected at the 1% significance level. Therefore, we can argue that the parameters of the model exhibit co-integrating relationships. This study cites urbanization, artificial intelligence innovation, foreign direct investment, gross domestic product, and energy consumption as the enduring driving forces. Furthermore, these factors necessitate the system's initial response to a typical stochastic disturbance. In summary, variations in these three parameters influence CO2 emissions in the United States.

Table 4. Results of ARDL bound test.

| Test Statistic | Value | Signif. | I(0) | I(1) |
|---|---|---|---|---|
| F-statistic | 5.3421 | 10% | 2.07 | 3 |
| k | 5 | 5% | 2.43 | 3.27 |
| | | 2.50% | 2.81 | 3.84 |
| | | 1% | 3.10 | 4.20 |

Table 4 and table 5 adopts the dynamic ARDL model to demonstrate the short- and long-term effects of LGDP, LAI, LENU, LFDI, and LURBA on $LCO_2$ in the USA. In terms of LGDP, a 1% boost in LGDP will increase the LCO2 by 0.028% in the long-term run and 0.012% in the short run. This suggests that economic expansion alone may notably contribute to environmental degradation in this setting, as GDP has a positive impact on the $CO_2$ emission level. A few studies have concluded that a boost in the GDP has a negative impact on the environment. This includes Ahmed et al. [106] in Japan; Raihan and Tuspekova [107] within Kazakhstan; Orhan et al. [108] on India; Ali et al. [109] in Malaysia; Shang et al.

[58] for ASEAN countries. However, Le and Ozturk [110], Zhan et al.[111], Tufail et al.[112] and He et al.[113] discovered the opposite outcome. They also concluded that economic pressures no longer adversely affect the natural world. Likewise, Awosusi et al. [114] demonstrated that there is no significant correlation between CO2 emissions and GDP in MINT economies.

On the other hand, the coefficients for LAI indicate a positive correlation with LCO$_2$, implying a 0.054% long-term fall and 0.076% short-term cut in LCO2 for every 1% rise in PAI. Thus, private investment in artificial intelligence in the United States significantly contributes to environmental sustainability. AI enhances energy efficiency and promotes environmental sustainability by decreasing carbon emissions. Multiple researchers like Nishant et al. [37], Chen et al.[31] and Zhao et al.[115] support this result, stating that AI technologies improve the environmental conditions in different regions. Moreover, Wang et al. [116] discovered that invention patents exhibited no significant correlation with emissions decrease. Conversely, LCO$_2$ is negatively associated with LENU in both the long and short run, and this relationship is statistically significant. These findings suggest that energy consumption has an adverse impact on the USA ecosystem. Specifically, a 1% increase in LENU increases LCO$_2$ by 0.617% in the long run and by 0.321% in the short run. The utilization of energy results in increased carbon emissions, as fossil fuels, the predominant energy source, emit substantial CO$_2$ during combustion. This result is consistent with the research of Islam et al. [117] in Bangladesh, Kim [118] in OECD countries, Nurgazina et al. [119] in Malaysia, Akbota and Baek [120] within Kazakhstan, and Odugbesan and Adebayo [121] in Nigeria. On the other hand, the findings of Namahoro et al. [122], Bhat [123], and Sharif et al. [4] concluded that energy use can have a negative impact on the environment by increasing the pollution level.

Table 5. Results of ARDL short-run and Long-run.

| Variable | Coefficient | Std. Error | t-Statistic | Prob. |
|---|---|---|---|---|
| Long-run Estimation | | | | |
| LGDP | 0.028 | 0.5432 | 0.0771 | 0.010 |
| LAI | -0.054 | 0.0140 | -2.3501 | 0.003 |
| LENU | 0.617 | 0.1177 | 1.4516 | 0.012 |
| LFDI | 0.018 | 0.0253 | 2.6532 | 0.014 |
| LURBA | 0.710 | 0.5406 | 2.0061 | 0.021 |
| C | 10.872 | 4.0321 | 3.0562 | 0.000 |
| Short-run Estimation | | | | |
| D(LCO2(-1)) | 0.517 | 0.1023 | 1.2054 | 0.041 |
| D(LGDP) | 0.012 | 0.3621 | 5.0452 | 0.000 |
| D(LAI) | -0.076 | 0.0054 | -3.1802 | 0.025 |
| D(LENU) | 0.321 | 0.1072 | 2.5638 | 0.024 |
| D(LFDI) | 0.031 | 0.1892 | -4.6723 | 0.000 |
| D(LURB) | 0.641 | 1.4912 | 2.6732 | 0.022 |
| CointEq(-1)* | -0.398 | 0.1040 | -4.4572 | 0.003 |

Standard errors in parentheses
*** p<0.01, ** p<0.05, * p<0.1

Similarly, there is an unfavorable correlation between LFDI and LCO$_2$, with each 1% increase in FDI increases the CO$_2$ emission by 0.018% in the long run and 0.031% in the short run. This result is significant at the 1% level. One possible reason is that FDI frequently results in elevated industrial activity, increased energy consumption, and intensified resource exploitation; hence, it amplifies environmental deterioration. Nie et al. [124], and Zhang et al. [125] corroborate this result. Conversely, Azam and Raza [126], Shah et al.[127], Pazienza [128] and Pradhan et al. [129] revealed that FDI can mitigate CO$_2$ emissions and improve biodiversity quality. Additionally, the positive and statistically significant URBA coefficients indicate that both long-term and short-term increases in LURBA negatively affect environmental quality. A 1% increase in URBA raises LCO$_2$ by 0.710% in the long run and by 0.641% in the short run. These findings suggest that the current urbanization structure in the United States is not conducive to reducing pollution. Several researchers have also observed a similar outcome, including Yuan et al. [130] in China, Ali et al. [131] in Pakistan, Anwar et al. [132] in Asian economies, Sikder et al.[133] in developing economies, and Raihan et al. [71]. However, studies conducted by Wang et al.[76], Acheampong [12], Zhu et al. [134], and Gasimli et al. [135] have demonstrated that urbanization enhances environmental sustainability by reducing carbon dioxide emissions.

The DOLS, FMOLS, and CCR techniques are supplementary methodologies utilized to evaluate the validity and reliability of the ARDL results. Table 6 outlines the robustness findings.

In the FMOLS model, the coefficients for LGDP are statistically significant at the 1% level and have positive values. A 1% increase in GDP causes the LCO$_2$ to rise by 0.245%. Additionally, a 1% increment in LAI leads to a 0.034% drop in CO$_2$ emission in the USA. Furthermore, a 1% boosts in LENU and LFDI and LURBA upsurges LCO$_2$ by 0.074%, 0.053% and 0.231%, respectively. It indicates that GDP, energy consumption, FDI and urbanization are not good for better for the ecosystem in USA. These findings corroborate the ARDL short and long-run estimation results, with LGDP, LAI, LENU and LFDI significant at the 1% level, while LURBA are significant at the 5% level. In the DOLS model, a 1% spike in LGDP, LENU, LFDI and LURBA results in an average rise of 0.316%, 0.023%, 0.039%, and 0.143% in LCO$_2$, respectively. Similar to the ARDL findings, a 1% rise in LAI leads to a 0.010% reduction in LCO$_2$, and the coefficient for LAI is significant at the 5% level.

In the CCR model, a 1% increase in LGDP, LENU, LFDI and LURBA leading to an average rise of 0.217%, 0.354%, 0.049%, and 0.205% in LCO$_2$, respectively. However, a 1% increase in LAI causes an average 0.037% decrease in LCO$_2$, confirming the ARDL results except for the LAI case. In this case all the factors are significant at 1% level, while LURBA is significant at 5% thresholds. These robustness checks confirm that the ARDL model's findings are reliable, as evidenced by the statistically significant values across FMOLS, DOLS, and CCR computations.

Table 6. Results of Robustness check.

| Variables | FMOLS | DOLS | CCR |
| --- | --- | --- | --- |
| LGDP | 0.245*** | 0.316*** | 0.217*** |
| LAI | -0.034*** | -0.010** | -0.037*** |
| LENU | 0.074*** | 0.023*** | 0.354*** |
| LFDI | 0.052*** | 0.039** | 0.049*** |
| LURB | 0.231** | 0.143** | 0.205** |
| C | 10.342*** | 10.052*** | 10.034*** |

Standard errors in parentheses
*** p<0.01, ** p<0.05, * p<0.1

Table 7 presents the conclusions of causal relationships among several economic indices. An F-statistic of 3.38423 and a p-value of 0.0102 suggest that LLGDP does not cause Granger-cause $LCO_2$. This indicates the rejection of the null hypothesis asserting no correlation between the variables at the 1% significance level. Also, p-values below the usual level of significance supported the finding that LAI, LENU, LFDI and LURBA all had a single-direction effect on $LCO_2$. Consequently, under these conditions, we dismiss the null hypothesis, asserting the absence of causation. On the other hand, p-values higher than the usual significance level showed that there was no significant causal relationship from $LCO_2$ to LGDP, LAI, LENU, LFDI and LURBA. The null hypothesis, which posits the absence of causality in these interactions, is not effectively disproved.

Table 7. Results of Pairwise Granger Causality test

| Null Hypothesis | Obs | F-Statistic | Prob. |
| --- | --- | --- | --- |
| LGDP $\neq$ LCO2 | 30 | 3.8423 | 0.0102 |
| LCO2 $\neq$ LGDP |  | 0.7345 | 0.1203 |
| LAI $\neq$ LCO2 | 30 | 3.0201 | 0.0073 |
| LCO2 $\neq$ LAI |  | 0.0123 | 0.8341 |
| LENU $\neq$ LCO2 | 30 | 4.6512 | 0.0143 |
| LCO2 $\neq$ LENU |  | 0.6712 | 0.7612 |
| LFDI $\neq$ LCO2 | 30 | 4.8061 | 0.0041 |
| LCO2 $\neq$ LFDI |  | 0.7623 | 0.0871 |
| LURB $\neq$ LCO2 | 30 | 4.7032 | 0.0191 |
| LCO2 $\neq$ LURB |  | 0.4121 | 0.5412 |

The results of the diagnostic evaluation are shown in Table 8. The findings indicated that the efficacy of all diagnostic techniques is negligible, and the null hypothesis remains indisputable. The Jarque-Bera test, yielding a p-value of 0.4321, suggests a normal distribution of the residuals. The Lagrange multiplier analysis indicates the absence of serial correlation in the residuals, yielding a p-value of 0.1021. Finally, the Breusch-Pagan-Godfrey test indicates that the residuals do not display heteroscedasticity, yielding a p-value of 0.1283.

Table 8. The results of diagnostic tests

| Diagnostic tests | Coefficient | p-value | Decision |
| --- | --- | --- | --- |
| Jarque-Bera test | 1.2034 | 0.4321 | Residuals are normally distributed |
| Lagrange Multiplier test | 1.0982 | 0.1021 | No serial correlation exits |
| Breusch-Pagan-Godfrey test | 0.0452 | 0.1283 | No heteroscedasticity exists |

Furthermore, we use the CUSUM and CUSUM-SQ statistics to seek structural stability in residuals over long and short periods. The CUSUM-SQ plot stays on the critical line, as indicated in the figure 1, indicating that the results are within the critical limits. This shows that the parameters are satisfactory and consistent at the 5% level of significance.

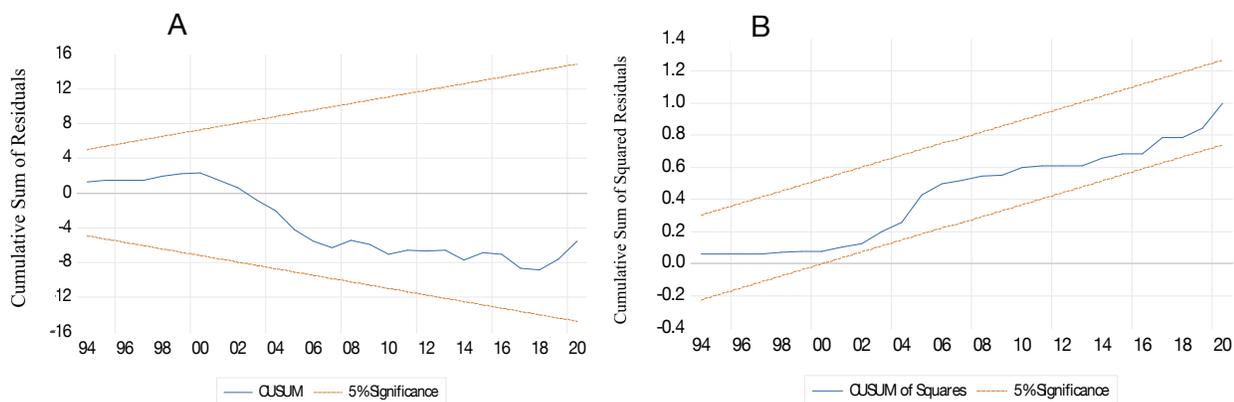

Figure 1. (A) CUSUM test for structural stability in residuals, (B) CUSUM of squares test for structural stability in residuals.

## Conclusion and Policy Implications

This paper thoroughly examines the impact of AI innovation, economic growth, foreign direct investment, energy consumption, and urbanization on $CO_2$ emissions in the USA from 1990 to 2022. Combining the ARDL framework with the STIRPAT structure, the authors found that while AI innovation reduces stress on the environment, economic growth, energy use, foreign direct investment, and urbanization make these problems worse. The results of the ADF, PP, and DF-GLS tests show that the variables have different levels of integration and no unit root problems. The ARDL boundaries test offers additional evidence of co-integration, signifying solid long-term interactions. The ARDL findings show a positive correlation between AI innovation and $CO_2$ emissions in the USA, suggesting that AI technologies improve environmental health as long as appropriate environmental safeguards are in place. In contrast, the adverse correlations among GDP, ENU, FDI, URBA, and $CO_2$ emissions indicate that these elements lead to detrimental environmental consequences. Significant technological advancements, including energy conservation, sustainable foreign direct investment, and urban planning, may stimulate new concepts and the implementation of environmentally friendly practices by enhancing competitiveness and facilitating access to advanced technologies. Robustness checks using FMOLS, DOLS, and CCR enhance the credibility of the ARDL results, thereby increasing their trustworthiness. Another thing is that Pairwise Granger causality tests show strong one-way links between $LCO_2$ and LGDP, LAI, LENU, LFDI, and LURBA. These linkages highlight the significant impact of economic transformations, private investments in AI, and advancements in green energy utilization on ecological sustainability dynamics in the USA. The research suggests various policy measures to promote sustainable economic growth in the United States, including the use of foreign direct investment, technological advancements, the implementation of green energy, and the development of sustainable urban infrastructure.

The findings emphasize the need for targeted policies to balance economic growth with environmental sustainability. Promoting AI-driven technologies and green energy initiatives can reduce $CO_2$ emissions while fostering innovation. Policies should prioritize sustainable foreign direct investment (FDI) by incentivizing eco-friendly projects and encouraging energy-efficient practices. Urban planning reforms must focus on developing smart, sustainable cities to mitigate the environmental impacts of urbanization. Additionally, integrating AI in energy management and industrial processes can enhance efficiency and reduce environmental stress. These measures collectively support sustainable economic growth while safeguarding ecological health in the USA.

## Declaration

**Ethics approval/declaration:** Not applicable.
**Consent to participate:** Not applicable.
**Consent for publication:** Not applicable.
**Acknowledgment:** Not applicable.
**Conflict of interest:** The authors declare no conflict of interest.
**Data availability:** Data will be available upon reasonable request from corresponding author.
**Authors contribution:** Azizul Hakim Rafi, Abdullah Al Abrar Chowdhury, and Adita Sultana from contributed equally to the conceptualization, data collection, and analysis of the study. Abdulla All Noman from provided support in data interpretation and contributed to the manuscript's critical revision. All authors reviewed and approved the final manuscript.

## References

1. Raihan, A., Voumik, L. C., Ridwan, M., Ridzuan, A. R., Jaaffar, A. H., & Yusoff, N. Y. M. (2023). From growth to green: navigating the complexities of economic development, energy sources, health spending, and carbon emissions in Malaysia. Energy Reports, 10, 4318-4331.
2. Voumik, L. C., Rahman, M. H., Rahman, M. M., Ridwan, M., Akter, S., & Raihan, A. (2023). Toward a sustainable future: Examining the interconnectedness among Foreign Direct Investment (FDI), urbanization, trade openness, economic growth, and energy usage in Australia. Regional Sustainability, 4(4), 405-415.
3. Pattak, D. C., Tahrim, F., Salehi, M., Voumik, L. C., Akter, S., Ridwan, M., ... & Zimon, G. (2023). The driving factors of Italy's $CO_2$ emissions based on the STIRPAT model: ARDL, FMOLS, DOLS, and CCR approaches. Energies, 16(15), 5845.
4. Raihan, A., Tanchangya, T., Rahman, J., & Ridwan, M. (2024). The Influence of Agriculture, Renewable Energy, International Trade, and Economic Growth on India's Environmental Sustainability. Journal of Environmental and Energy Economics, 37-53.
5. Ahmad, S., Raihan, A., & Ridwan, M. (2024). Role of economy, technology, and renewable energy toward carbon neutrality in China. Journal of Economy and Technology.
6. Ridwan, M., Raihan, A., Ahmad, S., Karmakar, S., & Paul, P. (2023). Environmental sustainability in France: The role of alternative and nuclear energy, natural resources, and government spending. Journal of Environmental and Energy Economics, 2(2), 1-16.
7. Raihan, A., Ridwan, M., Tanchangya, T., Rahman, J., & Ahmad, S. (2023). Environmental Effects of China's Nuclear Energy within the Framework of Environmental Kuznets Curve and Pollution Haven Hypothesis. Journal of Environmental and Energy Economics, 2(1), 1-12.
8. Raihan, A., Hasan, M. A., Voumik, L. C., Pattak, D. C., Akter, S., & Ridwan, M. (2024). Sustainability in Vietnam: Examining Economic Growth, Energy, Innovation, Agriculture, and Forests' Impact on $CO_2$ Emissions. World Development Sustainability, 100164.
9. Ridwan, M., Urbee, A. J., Voumik, L. C., Das, M. K., Rashid, M., & Esquivias, M. A. (2024). Investigating the environmental Kuznets curve hypothesis with urbanization, industrialization, and service sector for six South Asian Countries: Fresh evidence from Driscoll Kraay standard error. Research in Globalization, 8, 100223.
10. Raihan, A., Bala, S., Akther, A., Ridwan, M., Eleais, M., & Chakma, P. (2024). Advancing environmental sustainability in the G-7: The impact of the digital economy, technological innovation, and financial accessibility using panel ARDL approach. Journal of Economy and Technology.

11. Raihan, A., Ridwan, M., & Rahman, M. S. (2024). An exploration of the latest developments, obstacles, and potential future pathways for climate-smart agriculture. Climate Smart Agriculture, 100020.
12. Urbee, A. J., Ridwan, M., & Raihan, A. (2024). Exploring Educational Attainment among Individuals with Physical Disabilities: A Case Study in Bangladesh. Journal of Integrated Social Sciences and Humanities.
13. Ridwan, M. (2023). Unveiling the powerhouse: Exploring the dynamic relationship between globalization, urbanization, and economic growth in Bangladesh through an innovative ARDL approach.
14. Islam, S., Raihan, A., Ridwan, M., Rahman, M. S., Paul, A., Karmakar, S., ... & Al Jubayed, A. (2023). The influences of financial development, economic growth, energy price, and foreign direct investment on renewable energy consumption in the BRICS. Journal of Environmental and Energy Economics, 2(2), 17-28.
15. Voumik, L. C., Akter, S., Ridwan, M., Ridzuan, A. R., Pujiati, A., Handayani, B. D., ... & Razak, M. I. M. (2023). Exploring the factors behind renewable energy consumption in Indonesia: Analyzing the impact of corruption and innovation using ARDL model. International Journal of Energy Economics and Policy, 13(5), 115-125.
16. Ridzuan, A. R., Rahman, N. H. A., Singh, K. S. J., Borhan, H., Ridwan, M., Voumik, L. C., & Ali, M. (2023, May). Assessing the Impact of Technology Advancement and Foreign Direct Investment on Energy Utilization in Malaysia: An Empirical Exploration with Boundary Estimation. In International Conference on Business and Technology (pp. 1-12). Cham: Springer Nature Switzerland.
17. Rahman, J., Raihan, A., Tanchangya, T., & Ridwan, M. (2024). Optimizing the digital marketing landscape: A comprehensive exploration of artificial intelligence (AI) technologies, applications, advantages, and challenges. Frontiers of Finance, 2(2).
18. Ahmad, S., Raihan, A., & Ridwan, M. (2024). Pakistan's trade relations with BRICS countries: trends, export-import intensity, and comparative advantage. Frontiers of Finance, 2(2).
19. Raihan, A., Rahman, J., Tanchangtya, T., Ridwan, M., & Islam, S. (2024). An overview of the recent development and prospects of renewable energy in Italy. Renewable and Sustainable Energy, 2(2), 0008.
20. Tanchangya, T., Raihan, A., Rahman, J., Ridwan, M., & Islam, N. (2024). A bibliometric analysis of the relationship between corporate social responsibility (CSR) and firm performance in Bangladesh. Frontiers of Finance, 2(2).
21. Raihan, A., Rahman, J., Tanchangya, T., Ridwan, M., & Bari, A. B. M. (2024). Influences of economy, energy, finance, and natural resources on carbon emissions in Bangladesh. Carbon Research, 3(1), 1-16.
22. Onwe, J. C., Ridzuan, A. R., Uche, E., Ray, S., Ridwan, M., & Razi, U. (2024). Greening Japan: Harnessing energy efficiency and waste reduction for environmental progress. Sustainable Futures, 8, 100302.
23. Ridwan, M., Akther, A., Al Absy, M. S. M., Tahsin, M. S., Ridzuan, A. R., Yagis, O., & Mukhtar, K. J. (2024). The Role of Tourism, Technological Innovation, and Globalization in Driving Energy Demand in Major Tourist Regions. International Journal of Energy Economics and Policy, 14(6), 675-689.
24. Ridwan, M., Aspy, N. N., Bala, S., Hossain, M. E., Akther, A., Eleais, M., & Esquivias, M. A. (2024). Determinants of environmental sustainability in the United States: analyzing the role of financial development and stock market capitalization using LCC framework. Discover Sustainability, 5(1), 319.

25. Islam, S., Raihan, A., Paul, A., Ridwan, M., Rahman, M. S., Rahman, J., ... & Al Jubayed, A. (2024). Dynamic Impacts of Sustainable Energies, Technological Innovation, Economic Growth, and Financial Globalization on Load Capacity Factor in the Top Nuclear Energy-Consuming Countries. Journal of Environmental and Energy Economics, 1-14.
26. Raihan, A., Voumik, L. C., Ridwan, M., Akter, S., Ridzuan, A. R., Wahjoedi, ... & Ismail, N. A. (2024). Indonesia's Path to Sustainability: Exploring the Intersections of Ecological Footprint, Technology, Global Trade, Financial Development and Renewable Energy. In Opportunities and Risks in AI for Business Development: Volume 1 (pp. 1-13). Cham: Springer Nature Switzerland.
27. Manu SB, Sulaiman C (2017) Environmental Kuznets curve and the relationship between energy consumption, economic growth and CO2 emissions in Malaysia. J Econ Sustain Dev 8(16):142–148
28. Saudi, M. H. M., Sinaga, O., & Jabarullah, N. H. (2019). The role of renewable, non-renewable energy consumption and technology innovation in testing environmental Kuznets curve in Malaysia. International Journal of Energy Economics and Policy, 9(1), 299–307.
29. Zubair, A. O., Samad, A. R. A., & Dankumo, A. M. (2020). Does gross domestic income, trade integration, FDI inflows, GDP, and capital reduce CO2 emissions? An empirical evidence from Nigeria. Current Research in Environmental Sustainability, 2, 100009. https://doi.org/10.1016/j.crsust.2020.100009
30. Sarkar MSK, Al-Amin AQ, Mustapa SI, Ahsan MR (2019) Energy consumption, CO2 emission and economic growth: empirical evidence for Malaysia. Int J Environ Sustain Dev 18(3):318–334
31. Chen F, Wang L, Gu Q, Wang M, Ding X (2022) Nexus between natural resources, financial development, green innovation and environmental sustainability in China: Fresh insight from novel quantile ARDL. Resour Policy 79:102955. https://doi.org/10.1016/j.resourpol.2022.102955
32. Etokakpan MU, Solarin SA, Yorucu V, Bekun FV, Sarkodie SA (2020) Modeling natural gas consumption, capital formation, globalization, CO2 emissions and economic growth nexus in Malaysia: Fresh evidence from combined co-integration and causality analysis. Energy Strategy Rev 31:100526
33. Wada, I., Faizulayev, A., & Bekun, F. V. (2021). Exploring the role of conventional energy consumption on environmental quality in Brazil: Evidence from cointegration and conditional causality. Gondwana Research, 98, 244-256
34. Rjoub, H., Odugbesan, J. A., Adebayo, T. S., & Wong, W.-K. (2021). Sustainability of the moderating role of financial development in the determinants of environmental degradation: Evidence from Turkey. Sustainability, 13(4), 1844. https://doi.org/10.3390/SU13041844
35. He, X., Adebayo, T. S., Kirikkaleli, D., & Umar, M. (2021). Consumption-based carbon emissions in Mexico: An analysis using the dual adjustment approach. Sustainable Production and Consumption, 27, 947–957. https://doi.org/10.1016/J.SPC.2021.02.020
36. Muhammad, S., Long, X., Salman, M., & Dauda, L. (2020). Effect of urbanization and international trade on CO2 emissions across 65 belt and road initiative countries. Energy, 196, 117102. https://doi.org/10.1016/j.energy.2020.117102
37. Nishant, R., Kennedy, M., & Corbett, J. (2020). Artificial intelligence for sustainability: Challenges, opportunities, and a research agenda. International Journal of Information Management, 53, 102104. https://doi.org/10.1016/j.ijinfomgt.2020.102104
38. Kaligambe, A., Fujita, G., & Keisuke, T. (2022). Estimation of Unmeasured Room Temperature, Relative Humidity, and CO2 Concentrations for a Smart Building Using Machine Learning and Exploratory Data Analysis. Energies, 15(12), 4213.

*Journal of Environmental and Energy Economics*

39. Asadnia, M., Khorasani, A. M., & Warkiani, M. E. (2017). An Accurate PSO-GA Based Neural Network to Model Growth of Carbon Nanotubes. Journal of Nanomaterials, 2017(1), 9702384. https://doi.org/10.1155/2017/9702384
40. Liu, L., Yang, K., Fujii, H., & Liu, J. (2021). Artificial intelligence and energy intensity in China's industrial sector: Effect and transmission channel. Economic Analysis and Policy, 70, 276-293.
41. Chen, P., Gao, J., Ji, Z., Liang, H., & Peng, Y. (2022). Do artificial intelligence applications affect carbon emission performance?—Evidence from panel data analysis of Chinese cities. Energies, 15(15), 5730.
42. Negi, R. (2018). Global investment scenario of Artificial Intelligence (AI): A study with reference to China, India, and United States. Muthusamy, A., & Negi.
43. Pachot, A., & Patissier, C. (2022). Towards sustainable artificial intelligence: An overview of environmental protection uses and issues. arXiv preprint arXiv:2212.11738. https://doi.org/10.47852/bonviewGLCE3202608
44. Aktar, M. A., Alam, M. M., & Al-Amin, A. Q. (2021). Global economic crisis, energy use, $CO_2$ emissions, and policy roadmap amid COVID-19. Sustainable Production and Consumption, 26, 770-781.
45. Li Q, Cherian J, Shabbir MS, Sial MS, Li J, Mester I, Badulescu A (2021) Exploring the relationship between renewable energy sources and economic growth. The Case of SAARC Countries. Energies 14(3):520
46. Raihan, A., Begum, R.A., Nizam, M. et al. Dynamic impacts of energy use, agricultural land expansion, and deforestation on $CO_2$ emissions in Malaysia. Environ Ecol Stat 29, 477–507 (2022c). https://doi.org/10.1007/s10651-022-00532-9
47. Adebayo TS, Kalmaz DB (2021) Determinants of $CO_2$ emissions: Empirical evidence from Egypt. Environ Ecol Stat 28:239–262. https://doi.org/10.1007/s10651-020-00482-0
48. Adebayo TS, Awosusi AA, Adeshola I (2020) Determinants of $CO_2$ Emissions in Emerging Markets: An Empirical Evidence from MINT Economies. Int J Renew Energy Dev 9(3):411–422. https://doi.org/10.14710/ijred.2020.31321
49. Raihan, A., & Tuspekova, A. (2022). The nexus between economic growth, renewable energy use, agricultural land expansion, and carbon emissions: New insights from Peru. Energy Nexus, 6, 100067.
50. Abbasi, K.R., Adedoyin, F.F. RETRACTED ARTICLE: Do energy use and economic policy uncertainty affect $CO_2$ emissions in China? Empirical evidence from the dynamic ARDL simulation approach. Environ Sci Pollut Res 28, 23323–23335 (2021). https://doi.org/10.1007/s11356-020-12217-6
51. Balsalobre-Lorente, D., Driha, O. M., Halkos, G., & Mishra, S. (2022). Influence of growth and urbanization on $CO_2$ emissions: The moderating effect of foreign direct investment on energy use in BRICS. Sustainable Development, 30(1), 227-240. https://doi.org/10.1002/sd.2240
52. Saqib N (2021) Energy consumption and economic growth: empirical evidence from MENA region. Int J Energy Econ Policy 11(6):191–197. https://doi.org/10.32479/ijeep.11931
53. Younis I, Naz A, Shah SAA, Nadeem M, Longsheng C (2021) Impact of stock market, renewable energy consumption and urbanization on environmental degradation: new evidence from BRICS countries. Environ Sci Pollut Res 28:31549–31565. https://doi.org/10.1007/s11356-021-12731-1
54. Salari, M., Javid, R. J., & Noghanibehambari, H. (2021). The nexus between $CO_2$ emissions, energy consumption, and economic growth in the US. Economic Analysis and Policy, 69, 182-194. https://doi.org/10.1016/j.eap.2020.12.007

Science Research Publishers 71

55. Pao HT, Tsai CM (2010) CO2 emissions, energy consumption and economic growth in BRIC countries. Energy Policy 38:7850–7860
56. Gao D, Li G, Li Y, Gao K (2022) Does FDI improve green total factor energy efficiency under heterogeneous environmental regulation? Evidence from China. Environ Sci Pollut Res 29:25665–25678. https://doi.org/10.1007/s11356-021-17771-1
57. Jafri, A.H., Abbas, S., Abbas, S.M.Y. et al. Caring for the environment: measuring the dynamic impact of remittances and FDI on CO2 emissions in China. Environ Sci Pollut Res 29, 9164–9172 (2022). https://doi.org/10.1007/s11356-021-16180-8
58. Shang Y, Razzaq A, Chupradit S et al (2022) The role of renewable energy consumption and health expenditures in improving load capacity factor in ASEAN countries: exploring new paradigm using advance panel models. Renew Energy 191:715–722. https://doi.org/10.1016/J.RENENE.2022.04.013
59. Xie Q, Wang X, Cong X (2020) How does foreign direct investment affect CO2 emissions in emerging countries? New findings from a nonlinear panel analysis. J Clean Prod 249:119422
60. Qin B, Gai Y, Ge L et al (2022) FDI, technology spillovers, and green innovation: theoretical analysis and evidence from China. Energies (Basel) 15:7497. https://doi.org/10.3390/en15207497
61. Gill AR, Hassan S, Haseeb M (2019) Moderating role of financial development in environmental Kuznets: a case study of Malaysia. Environ Sci Pollut Res 26(33):34468–34478
62. Lin, H., Wang, X., Bao, G., & Xiao, H. (2022). Heterogeneous spatial effects of FDI on CO2 emissions in China. Earth's Future, 10(1), e2021EF002331.
63. Eskeland GS, Harrison AE (2003) Moving to greener pastures? Multinationals and the pollution haven hypothesis. J Dev Econ 70(1):1–23. https://doi.org/10.1016/S0304-3878(02)00084-6
64. Wang, W. Z., Liu, L. C., Liao, H., & Wei, Y. M. (2021). Impacts of urbanization on carbon emissions: An empirical analysis from OECD countries. Energy Policy, 151, 112171. https://doi.org/10.1016/j.enpol.2021.112171
65. Pata, U. K., & Samour, A. (2022). Do renewable and nuclear energy enhance environmental quality in France? A new EKC approach with the load capacity factor. Progress in Nuclear Energy, 149, 104249.
66. Abbas A, Moosa I, Ramiah V (2022) The contribution of human capital to foreign direct investment inflows in developing countries. J Intellect Cap 23(1):9–26. https://doi.org/10.1108/JIC-12-2020-0388
67. Haug AA, Ucal M (2019) The role of trade and FDI for CO2 emissions in Turkey: nonlinear relationships. Energy Econ 81:297–307. https://doi.org/10.1016/j.eneco.2019.04.006
68. Nguyen HM, Nguyen LD (2018) The relationship between urbanization and economic growth: an empirical study on ASEAN countries. Int J Soc Econ. 45(2):316–339
69. Mahmood, H., Alkhateeb, T. T. Y., & Furqan, M. (2020). Industrialization, urbanization and CO2 emissions in Saudi Arabia: Asymmetry analysis. Energy Reports, 6, 1553-1560. https://doi.org/10.1016/j.egyr.2020.06.004
70. Parshall, L., Gurney, K., Hammer, S. A., Mendoza, D., Zhou, Y., & Geethakumar, S. (2010). Modeling energy consumption and CO2 emissions at the urban scale: Methodological challenges and insights from the United States. Energy Policy, 38(9), 4765–4782.
71. Raihan, A., Atasoy, F. G., Atasoy, M., Ridwan, M., & Paul, A. (2022b). The role of green energy, globalization, urbanization, and economic growth toward environmental sustainability in the United States. Journal of Environmental and Energy Economics, 1(2), 8-17. https://doi.org/10.56946/jeee.v1i2.377

72. Sufyanullah, K., Ahmad, K. A., & Ali, M. A. S. (2022). Does emission of carbon dioxide is impacted by urbanization? An empirical study of urbanization, energy consumption, economic growth and carbon emissions-Using ARDL bound testing approach. Energy Policy, 164, 112908.
73. Yao, F., Zhu, H., & Wang, M. (2021). The impact of multiple dimensions of urbanization on CO2 emissions: a spatial and threshold analysis of panel data on China's prefecture-level cities. Sustainable Cities and Society, 73, 103113. https://doi.org/10.1016/j.scs.2021.103113
74. Raihan, A., Tanchangya, T., Rahman, J., Ridwan, M., & Ahmad, S. (2022a). The influence of Information and Communication Technologies, Renewable Energies and Urbanization toward Environmental Sustainability in China. Journal of Environmental and Energy Economics, 1(1), 11-23. https://doi.org/10.56946/jeee.v1i1.351
75. Gierałtowska, U., Asyngier, R., Nakonieczny, J., & Salahodjaev, R. (2022). Renewable energy, urbanization, and CO2 emissions: a global test. Energies, 15(9), 3390.
76. Wang M, Zhang X, Hu Y (2021) The green spillover effect of the inward foreign direct investment: market versus innovation. J Clean Prod 328:129501. https://doi.org/10.1016/j.jclepro.2021.129501
77. Anwar, A., Younis, M., & Ullah, I. (2020). Impact of urbanization and economic growth on CO2 emission: a case of far east Asian countries. International Journal of Environmental Research and Public Health, 17(7), 2531.
78. Xu, D., Salem, S., Awosusi, A. A., Abdurakhmanova, G., Altuntaş, M., Oluwajana, D., ... & Ojekemi, O. (2022). Load capacity factor and financial globalization in Brazil: the role of renewable energy and urbanization. Frontiers in Environmental Science, 9, 823185. https://doi.org/10.3389/fenvs.2021.823185
79. Haseeb A, Xia E, Danish BMA, Abbas K (2018) Financial development, globalization, and CO2 emission in the presence of EKC: evidence from BRICS countries. Environ Sci Pollut Res 25(31):31283–31296. https://doi.org/10.1007/s11356-018-3034-7
80. Martínez, C. I. P., Piña, W. H. A., & Moreno, S. F. (2018). Prevention, mitigation and adaptation to climate change from perspectives of urban population in an emerging economy. Journal of Cleaner Production, 178, 314-324.
81. Diputra EM, Baek J (2018) Does growth good or bad for the environment in Indonesia? Int J Energy Econ Policy. 8(1):1–4
82. Wang, P., Wu, W., Zhu, B., & Wei, Y. (2013). Examining the impact factors of energy-related CO2 emissions using the STIRPAT model in Guangdong Province, China. Applied energy, 106, 65-71.
83. Ehrlich PR, Holdren JP (1971) Impact of population growth. Science 171(3977):1212–1217. https://doi.org/10.1126/science.171.3977.1212
84. Fan, Y., Liu, L. C., Wu, G., & Wei, Y. M. (2006). Analyzing impact factors of CO2 emissions using the STIRPAT model. Environmental impact assessment review, 26(4), 377-395.
85. Dietz T, Rosa EA (1994) Rethinking the environmental impacts of population. Affl Technol Human Ecol Rev 1(2):277–300
86. Sarker, B., & Khan, F. (2020). Nexus between foreign direct investment and economic growth in Bangladesh: an augmented autoregressive distributed lag bounds testing approach. Financial Innovation, 6(1), 10.
87. Elliott, G., Rothenberg, T. J., & Stock, J. H. (1992). Efficient tests for an autoregressive unit root.
88. Phillips, P. C., & Perron, P. (1988). Testing for a unit root in time series regression. Biometrics, 75(2), 335-346.
89. Dickey, D. A., & Fuller, W. A. (1979). Distribution of the estimators for autoregressive time series with a unit root. Journal of the American statistical association, 74(366a), 427-431.

90. Fuller, W. A. (2009). Introduction to statistical time series. John Wiley & Sons.
91. Vougas DV (2007) GLS detrending and unit root testing. Econ Lett 97:222–229
92. Sarkodie, S. A., & Owusu, P. A. (2020). How to apply the novel dynamic ARDL simulations (dynardl) and Kernel-based regularized least squares (krls). MethodsX, 7, 101160.
93. Pesaran, M. H., Shin, Y., & Smith, R. J. (2001). Bounds testing approaches to the analysis of level relationships. Journal of applied econometrics, 16(3), 289-326. https://doi.org/10.1002/jae.616
94. Chandio, A. A., Jiang, Y., Rauf, A., Ahmad, F., Amin, W., & Shehzad, K. (2020). Assessment of formal credit and climate change impact on agricultural production in Pakistan: A time series ARDL modeling approach. Sustainability, 12(13), 5241.
95. Murad, S.M.W., Salim, R., & Kibria, M.G. Asymmetric Effects of Economic Policy Uncertainty on the Demand for Money in India. J. Quant. Econ. 19, 451–470 (2021). https://doi.org/10.1007/s40953-021-00235-1
96. Alemu, K. (2020). The impact of population growth on economic growth: evidence from Ethiopia using ARDL approach. Journal of Economics and Sustainable Development, 11(3), 19-33.
97. Engle, R. F., & Granger, C. W. (1987). Co-integration and error correction: representation, estimation, and testing. Econometrica: journal of the Econometric Society, 251-276.
98. Phillips, P. C., & Hansen, B. E. (1990). Statistical inference in instrumental variables regression with I (1) processes. The review of economic studies, 57(1), 99-125.
99. Merlin, M. L., & Chen, Y. (2021). Analysis of the factors affecting electricity consumption in DR Congo using fully modified ordinary least square (FMOLS), dynamic ordinary least square (DOLS) and canonical cointegrating regression (CCR) estimation approach. Energy, 232, 121025.
100. Pesaran, M. H. (1997). The role of economic theory in modelling the long run. The economic journal, 107(440), 178-191.
101. Park, J. Y. (1992). Canonical cointegrating regressions. Econometrica: Journal of the Econometric Society, 60(1), 119-143.
102. Granger, C. W. (1969). Investigating causal relations by econometric models and cross-spectral methods. Econometrica: Journal of the Econometric Society, 37(3), 424-438.
103. Jarque, C. M., & Bera, A. K. (1987). A test for normality of observations and regression residuals. International Statistical Review/Revue Internationale de Statistique, 163-172. https://doi.org/10.2307/1403192
104. Breusch, T. S., & Pagan, A. R. (1979). A simple test for heteroscedasticity and random coefficient variation. Econometrica: Journal of the Econometric Society, 1287-1294. https://doi.org/10.2307/1911963
105. Brown, R. L., Durbin, J., & Evans, J. M. (1975). Techniques for Testing the Constancy of Regression Relationships over Time. Journal of the Royal Statistical Society: Series B (Methodological), 37(2), 149–163.
106. Ahmed, Z., Cary, M., Ali, S., Murshed, M., Ullah, H., & Mahmood, H. (2022). Moving toward a green revolution in Japan: symmetric and asymmetric relationships among clean energy technology development investments, economic growth, and CO2 emissions. Energy & Environment, 33(7), 1417-1440.
107. Raihan, A., & Tuspekova, A. (2022). Dynamic impacts of economic growth, energy use, urbanization, agricultural productivity, and forested area on carbon emissions: New insights from Kazakhstan. World Development Sustainability, 1, 100019. https://doi.org/10.1016/j.wds.2022.100019
108. Orhan A, Adebayo TS, Genç SY, Kirikkaleli D (2021) Investigating the linkage between economic growth and environmental sustainability in India: do agriculture and trade openness matter? Sustainability 13(9):4753. https://doi.org/10.3390/su13094753

109. Ali, S. S. S., Razman, M. R., & Awang, A. (2020). The nexus of population, GDP growth, electricity generation, electricity consumption and carbon emissions output in Malaysia. International Journal of Energy Economics and Policy, 10(3), 84-89.
110. Le, H. P., & Ozturk, I. (2020). The impacts of globalization, financial development, government expenditures, and institutional quality on CO2 emissions in the presence of environmental Kuznets curve. Environmental Science and Pollution Research, 27(18), 22680–22697.
111. Zhan, Z., Ali, L., Sarwat, S., Godil, D. I., Dinca, G., & Anser, M. K. (2021). A step towards environmental mitigation: do tourism, renewable energy and institutions really matter? A QARDL approach. Science of the Total Environment, 778, 146209.
112. Tufail, M., Song, L., Adebayo, T. S., Kirikkaleli, D., & Khan, S. (2021). Do fiscal decentralization and natural resources rent curb carbon emissions? Evidence from developed countries. Environmental Science and Pollution Research, 28(35), 49179-49190.
113. He, P., Ya, Q., Chengfeng, L., Yuan, Y., & Xiao, C. (2021). Nexus between environmental tax, economic growth, energy consumption, and carbon dioxide emissions: evidence from China, Finland, and Malaysia based on a Panel-ARDL approach. Emerging Markets Finance and Trade, 57(3), 698-712.
114. Awosusi AA, Adebayo TS, Adeshola I (2020) Determinants of CO2 Emissions in Emerging Markets: An Empirical Evidence from MINT Economies. Int J Renew Energy Dev 9(3):411–422.
115. Zhao, J., Xi, X. I., Na, Q. I., Wang, S., Kadry, S. N., & Kumar, P. M. (2021). The technological innovation of hybrid and plug-in electric vehicles for environment carbon pollution control. Environmental Impact Assessment Review, 86, 106506.
116. Wang, Z., Yang, Z., Zhang, Y., & Yin, J. (2012). Energy technology patents–CO2 emissions nexus: An empirical analysis from China. Energy Policy, 42, 248–260.
117. Islam, M.M., Khan, M.K., Tareque, M. et al. Impact of globalization, foreign direct investment, and energy consumption on CO2 emissions in Bangladesh: Does institutional quality matter? Environ Sci Pollut Res 28, 48851–48871 (2021). https://doi.org/10.1007/s11356-021-13441-4
118. Kim, D. K. (2020). Energy consumption, economic growth, and environmental degradation in OECD countries. Asian Journal of Economics and Empirical Research, 7(2), 242-250.
119. Nurgazina Z, Ullah A, Ali U, Koondhar MA, Lu Q (2021) The impact of economic growth, energy consumption, trade openness, and financial development on carbon emissions: empirical evidence from Malaysia. Environ Sci Pollut Res 28(42):60195–60208.
120. Akbota A, Baek J (2018) The environmental consequences of growth: empirical evidence from the Republic of Kazakhstan. Economies 6(1):19. https://doi.org/10.3390/economies6010019.
121. Odugbesan JA, Adebayo TS (2020) The symmetrical and asymmetrical effects of foreign direct investment and financial development on carbon emission: evidence from Nigeria. SN Appl Sci 2(12):1–15. https://doi.org/10.1007/s42452-020-03817-5.
122. Namahoro JP, Wu Q, Zhou N, Xue S (2021) Impact of energy intensity, renewable energy, and economic growth on CO2 emissions: evidence from Africa across regions and income levels. Renew Sustain Energy Rev 147:111233.
123. Bhat JA (2018) Renewable and non-renewable energy consumption-impact on economic growth and CO2 emissions in five emerging market economies. Environ Sci Pollut Res 25(35):35515–35530.
124. Nie, Y., Liu, Q., Liu, R., Ren, D., Zhong, Y., & Yu, F. (2022). The threshold effect of FDI on CO2 emission in belt and road countries. International Journal of Environmental Research and Public Health, 19(6), 3523.

*Journal of Environmental and Energy Economics*
125. Zhang D, Ozturk I, Ullah S (2022) Institutional factors-environmental quality nexus in BRICS: a strategic pillar of governmental performance. Econ Res-Ekonomska Istraživanja:1–13. https://doi.org/10.1080/1331677X.2022.2037446.
126. Azam, M., & Raza, A. (2022). Does foreign direct investment limit trade-adjusted carbon emissions: Fresh evidence from global data. Environmental Science and Pollution Research, 29(25), 37827–37841.
127. Shah SAR, Naqvi SAA, Nasreen S, Abbas N (2021) Associating drivers of economic development with environmental degradation: fresh evidence from Western Asia and North African region. Ecol Indic 126:107638. https://doi.org/10.1016/j.ecolind.2021.107638.
128. Pazienza P (2019) The impact of FDI in the OECD manufacturing sector on CO2 emission: evidence and policy issues. Environ Impact Assess Rev 77(1):60–68. https://doi.org/10.1016/j.eiar.2019.04.002.
129. Pradhan, A.K., Sachan, A., Sahu, U.K. et al. Do foreign direct investment inflows affect environmental degradation in BRICS nations? Environ Sci Pollut Res 29, 690–701 (2022). https://doi.org/10.1007/s11356-021-15678-5.
130. Yuan, M., Yin, C., Sun, Y., & Chen, W. (2019). Examining the associations between urban built environment and noise pollution in high-density high-rise urban areas: A case study in Wuhan, China. Sustainable Cities and Society, 50, 101678. https://doi.org/10.1016/j.scs.2019.101678.
131. Ali, R., Bakhsh, K., & Yasin, M. A. (2019). Impact of urbanization on CO2 emissions in emerging economy: evidence from Pakistan. Sustainable Cities and Society, 48, 101553.
132. Anwar, A., Sinha, A., Sharif, A. et al. The nexus between urbanization, renewable energy consumption, financial development, and CO2 emissions: evidence from selected Asian countries. Environ Dev Sustain 24, 6556–6576 (2022). https://doi.org/10.1007/s10668-021-01716-2.
133. Sikder, M., Wang, C., Yao, X., Huai, X., Wu, L., KwameYeboah, F., ... & Dou, X. (2022). The integrated impact of GDP growth, industrialization, energy use, and urbanization on CO2 emissions in developing countries: evidence from the panel ARDL approach. Science of the Total Environment, 837, 155795.
134. Zhu H, Xia H, Guo Y, Peng C (2018) The heterogeneous effects of urbanization and income inequality on CO2 emissions in BRICS economies: evidence from panel quantile regression. Environ Sci Pollut Res 25:17176–17193.
135. Gasimli O, ul Haq I, Gamage SKN et al (2019) Energy, trade, urbanization and environmental degradation nexus in Sri Lanka: bounds testing approach. Energies 12:1–16. https://doi.org/10.3390/en12091655.
Science Research Publishers     76